\documentclass{PoS}

\usepackage{slashed,amsmath,amssymb}
\usepackage{bm}
\newcommand{\vc}[1]{\mbox{\boldmath$#1$}}

\newcommand{\ud}{\mathrm{d}}

\newcommand{\be}{\begin{equation}}
\newcommand{\ee}{\end{equation}}
\newcommand{\bea}{\begin{eqnarray}}
\newcommand{\eea}{\end{eqnarray}}

\newcommand{\m}{{\scriptscriptstyle -}} 
\newcommand{\p}{{\scriptscriptstyle +}}

\newcommand{\LCv}[1]{\mathsf{#1}}

\newcommand{\LCperp}{{\scriptscriptstyle \perp}}

\title{Pair production in laser fields: finite size effects}

\ShortTitle{Pair production in laser fields}

\author{Thomas Heinzl\\
        School of Computing and Mathematics, University of Plymouth\\
        E-mail: \email{t.heinzl@plymouth.ac.uk}}

\author{\speaker{Anton Ilderton} \\ 
        Department of Physics, Ume\aa\ University\\
        E-mail: \email{anton.ilderton@physics.umu.se}}

\author{Mattias Marklund\\
        Department of Physics, Ume\aa\ University\\
        E-mail: \email{mattias.marklund@physics.umu.se}}
        
\abstract{We discuss pair creation in a strong laser background. Using lightfront field theory, we show that all the physics is contained in the lightfront momentum transfer from the laser, and probe, to the produced pair. The dependence of this momentum transfer on the geometry of the laser leads to resonance and diffraction effects in pair production spectra. The lightfront approach naturally explains the interpretation of laser-stimulated pair production as a multi-photon process creating pairs of an effective mass.}

\FullConference{Light Cone 2010 - LC2010\\
		June 14-18, 2010\\
		Valencia, Spain}

\begin{document}

\section{Introduction}
Fifty years since their invention, lasers now reach intensities of $10^{22}$ W/cm$^2$, delivering electric fields strong enough to ionise atoms, in femtosecond duration pulses. The energy density of these pulses reaches $10^{10}$ J/cm$^3$. The next generation of laser facilities will reach even higher field strengths -- the intensity of the Vulcan laser at CLF in the UK will this year be increased by a factor of ten, and by 2015 the Extreme Light Infrastructure (ELI) project will house a 100PW laser capable of an intensity of $10^{24}$ W/cm$^2$ and a peak electric field of $10^{14}$ V/cm \cite{Heinzl:2008an}.

This last value is two orders of magnitude below the critical field strength of QED, at which the field becomes strong enough to pull electron-positron pairs out of the vacuum: this is the Sauter-Schwinger electric field strength $E_S \sim m_e \sim 10^{16}$ V/cm. The derivation of the Sauter-Schwinger field strength is based on the assumption of a constant, homogeneous electric field, whereas the fields of a laser are pulsed and rapidly oscillating, and there are indications that in such fields the Sauter-Schwinger limit is lowered \cite{Bulanov:2004de, Bulanov:2010ei}. The study of finite size effects, stemming from finite pulse geometry, has therefore received quite some attention in recent years.

These investigations have, mostly, focussed on vacuum pair production in electric fields which remain spatially homogenous but become time dependent, and have revealed a rich substructure in the spectrum of the produced pairs \cite{Hebenstreit:2009km,Dumlu:2010ua,MockenNew}. Here, we will extend these investigations to stimulated pair creation  \cite{Schutzhold:2008pz}. Since the effects of the magnetic field can become important at high frequencies \cite{Ruf:2009zz}, we include it by modelling our laser as a plane wave of {\it finite} temporal (longitudinal) extent.  The approach is based on strong-field QED, and we will see that our calculations are naturally suited to lightfront co-ordinates. We begin by reviewing established results concerning particles in plane waves, and then go on to study some effects of finite beam geometry.

\section{Plane waves and the electron mass shift}
A laser pulse contains a huge number of photons and will therefore be treated as a classical, fixed background. We model the laser fields with a plane wave, i.e.\  a function of $k.x$ such that $k^2=0$, lightlike. We take $k.x\sim x^+$, lightfront time. The plane wave is
\be\label{F}
	F_{\mu\nu} = \dot{f}_j(k.x)  ( k_\mu a^j_\nu -a^j_\mu k_\nu ) \;,
\ee
with $j$ summed over 1 and 2, and a dot is differentiation with respect to $k.x$ (for later convenience). The polarisation vectors obey $k.a_i=0$ and $a^i.a^j=-m^2a^2/e^2\delta^{ij}$, defining an invariant, dimensionless amplitude $a$. The electric and magnetic fields are always orthogonal and of equal magnitude, so both Lorentz invariants constructed from the field vanish identically. One may construct an invariant $\mathcal T$ by employing the momentum vector $p_\mu$ of a probe electron moving in the field \cite{Heinzl:2008rh}:
\be
	\mathcal{T} = \frac{e^2}{m^2}\frac{p.T.p}{k.p^2} \;,
\ee
where $T_{\mu\nu}$ is the energy momentum tensor for $F_{\mu\nu}$. One may check that $\mathcal T$ reduces to the electromagnetic energy density seen by the electron in its instantaneous rest frame. This leads us to the question `what is $p_\mu$?' Classically, this is found by solving the  Lorentz force equation, 
\be\label{Lorentz}
	\frac{\ud}{\ud\tau}p_\mu = \frac{e}{m} F_{\mu\nu} p^\nu \;,
\ee
which is possible here because $k.p$ is conserved and therefore proper time $\tau$ may be identified with {\it lightfront} time $k.x\sim x^+$. Using this, the momentum and trajectory are easily calculated.  In the context of pair production, only plane waves which are of infinite extent in $k.x$, and are purely oscillatory, have until recently been thoroughly studied in the literature \cite{Nikishov:1963,Nikishov:1964a,Narozhnyi:1964}, so let us specialise to this `infinite plane wave' (IPW) case. Take linear polarisation, i.e.\ $f_1(k.x)=\sin{k.x}$, $f_2(k.x)=0$. As the fields oscillate, an electron undergoes rapid `quiver' motion, but the important quantities turn out to be {\it averages} taken over a cycle of the laser.  Taking the cycle-average of the momentum gives the `quasi-momentum' $q_\mu$, which, in terms of an initial momentum $p_\mu$ is
\be \label{QUASIMOM.INF}
  q_\mu = p_\mu + \frac{m^2 a_0^2}{2 k.p} \, k_\mu \; .
\ee
This is parameterised by the invariant $\mathcal T$ averaged over a cycle, written $a_0^2$, and which equals $a^2/2$ ($a^2$) for linear (circular)  polarisation.  $a_0$ parameterises all our intensity effects and is called the `dimensionless laser intensity parameter'. It is of $\mathcal{O}(10)$ for today's lasers and will reach at least $\mathcal{O}(10^3)$ at ELI. Squaring (\ref{QUASIMOM.INF}) one finds the celebrated electron mass shift,  $q^2 = m^2 (1 + a_0^2) =:m_*^2$ \cite{Sengupta:1952}. The relevance of the effective mass  is seen when one calculates scattering amplitudes. We focus on pair production stimulated by a probe photon, momentum $k'_\mu$, fired into the laser: this probe is required because, unlike a pure electric field, a single plane wave cannot create pairs from the vacuum, a consequence of both its Lorentz invariants vanishing\footnote{This may also be easily seen from momentum conservation. Can one relate these statements to the triviality of the lightfront vacuum? \cite{Heinzl:2003jy}.} \cite{Schwinger:1951nm}).

Sparing the details, one finds that in an IPW the S--matrix has the form \cite{Nikishov:1963,Nikishov:1964a,Narozhnyi:1964}:
\be\label{SSq}
	\frac{|S_{fi}|^2}{VT} = \sum\limits_{n> n_0^*} |M_n|^2 \delta^4( k'_\mu + n k_\mu= q_\mu + q'_\mu ) \;.
\ee
This {\it incoherent} sum is supported on the conservation of quasi-momentum; each term is therefore naturally associated with the effective process
\be
	\gamma(k') + n \gamma(k) \to e^-(q) + e^+(q') \;.
\ee
 in which the probe photon, together with $n$ photons of momentum $k_\mu$, referred to as `laser photons', create an electron-positron pair of rest mass $m_*$.  Indeed, the lower limit of the sum in (\ref{SSq}) is $n_0^* = 2m_*^2/k.k'$ (found from squaring the quasi-momentum relations), which just says that the number of photons taken from the laser should be sufficient to produce a {\it heavy} pair.

The bulk of intuition regarding laser-matter interactions is based on the above (and related) results: scattering processes are summed over the number of photons taken from the laser, and the electron acquires an effective mass $m_*$. Indeed, these results were used to analyse the SLAC E144 experiment which produced pairs by colliding the SLAC beam against a low intensity laser \cite{Bamber:1999zt}.  Although confirming nonlinear, or multi-photon, effects, the experiment did not offer an unambiguous detection of the effective mass \cite{McD-talk}. 

There are some loose ends to be tied up here: what does the `laser photon' number mean, when the background is classical? Is the prevalence of the effective mass over the rest mass due only to the fact that the fermions never escape the background field?  These issues may be addressed by incorporating the simplest of finite size effects: those due to finite pulse duration, which in our case means finite extent in lightfront time.  Regarding experiment, the assumption of a periodic plane wave is now hard to justify since a modern laser pulse has femtosecond duration, and contains an order of 10 cycles of the beam, around 1000 times fewer than in the laser used for the SLAC experiment. So let us now consider a `pulsed' plane wave and see what becomes of the mass shift and photon number\footnote{Calculating the pair production rate in a genuine `pulse-shaped' background in which the fields have finite extent in all directions is challenging, to say the least. See \cite{Heinzl:2009nd} for a  recent discussion in the context of nonlinear Compton scattering, i.e.\  the crossed process of pair production.}.  
\section{A lightfront approach to finite size effects}
\noindent\begin{tabular}{lc}
\begin{minipage}{0.4\textwidth}
\includegraphics[width=\columnwidth]{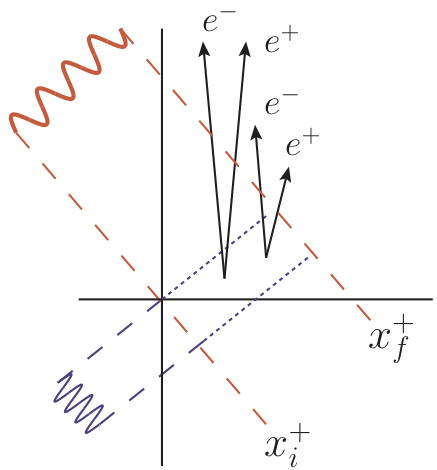}
\end{minipage} &
\noindent\begin{minipage}{0.56\textwidth}
Our plane wave is a function of $k.x\sim x^+$, or lightfront time. A second laser, or a probe photon, depends on another lightlike direction, say $x^-$ for a head-on collision. The setup is shown to the left: pairs are produced where the fields overlap, and the pairs escape the fields after a finite time. It is really {\it lightfront} time which is relevant here: front times bound the production region, we will see below that lightfront momentum fractions are the important objects in play, and the number of pairs created could be calculated from expectation values of the number operator at front times $x^+_f$ and $x^+_i$ (c.f.\  the calculation in \cite{Hallin:1994ad} for electric fields in the instant form). So, laser physics really is lightfront physics.
\end{minipage}
\end{tabular} \\ \newline
We now proceed to the scattering amplitude. The calculation is performed in the Furry picture \cite{MoortgatPick:2009zz}: the laser field is treated classically and preferably exactly (since $F_{\mu\nu}$ is characterised by $a_0>1$, a perturbative expansion in the field strength is not permissible) while scattered fermions and photons are quantised and treated in perturbation theory as normal.  Again, the most complete treatment is possible for plane waves. One begins by solving the Dirac equation in the background (\ref{F}), for which we need to choose a gauge. It is convenient to work in lightfront gauge, so $k.A\sim A^+=0$. The gauge potential for the field (\ref{F}) is then
\be
	A_\mu (k.x) = f_j(k.x) \, a^j_\mu \;,
\ee
and the `Volkov electron' solution to the Dirac equation is \cite{volkov:1935}
\be \label{VOLKOV}
  \Psi^{\,e^-}_p = \exp\bigg[-ip.x + \frac{i}{2k.p} \int\limits_{k.x}^{\infty} \! 2e A.p - e^2 A^2 \; ,\bigg]\bigg( 1 + \frac{e}{2k.p} \slashed{k} \slashed{A} \bigg) u_p \equiv e^{-ip.x -i I_p} \Gamma_{\!p}\, u_p \; ,
\ee
where $p$ is on-shell and $u_p$ is a free Dirac spinor.  The (tree level) amplitude for stimulated pair production of an electron with momentum $p$ and a positron with momentum $p'$ can be written in terms of the Volkov wavefunctions as follows 
\be \label{SFI}
	S_{fi} = -ie \int\! \ud^4x\   \overline \Psi^{\,e^-}_p(x) \slashed{\epsilon}\, e^{-ik'.x}\,  \Psi^{\,e^+}_{p'}(x) \;,
\ee
where $k'_\mu$ and $\epsilon$ are the momentum and polarisation vectors of the probe photon.  The problem at hand is essentially one dimensional: going to lightfront co-ordinates, the scattering amplitude reduces to a lightfront Fourier integral over $x^+$, or, for convenience, $k.x$,
\be \label{FOURIER.MFRAC}
  S_{fi} \!= \!\frac{1}{k_\p}\delta^3 (\bm{\mathsf{p}}' + \bm{\LCv{p}} -\bm{\LCv{k}}')\! \int\! \ud (k.x) \, e^{i(\LCv{y}+\LCv{y}'-\LCv{x}')k.x}{\mathcal{M}(k.x)}\;,
\ee
of the function $\mathcal{M}=\bar{u}_p \overline\Gamma_{p}\,\slashed{\varepsilon}\, \Gamma_{\m p'}v_{p'}\ e^{i(I_p-I_{\m p'})}$, with the lightfront three momenta $\bm{\LCv{p}} := (p_\m, \vc{p}_\LCperp)$ conserved, as follows from (\ref{VOLKOV}) being an eigenfunction of $\bm{\LCv{p}}$. Here we have introduced three boost invariant lightfront (longitudinal) momentum fractions, $\LCv{y}$, $\LCv{y}'$ and $\LCv{x}'$ which appear through the sum
\be\label{snew}
	 \LCv{y}+\LCv{y}' -\LCv{x}' \equiv  \frac{p_\p}{k_\p} +  \frac{p'_\p}{k_\p} -  \frac{k'_\p}{k_\p}\;,
\ee
and which we note are Fourier conjugate to the invariant phase $k.x$. To progress, one must choose an explicit $F_{\mu\nu}$. We are interested in the case of $F_{\mu\nu}$ having compact support $P$ in $k.x$. 

\subsubsection*{Example 1: a finite wavetrain and the IPW limit.}
We begin with a simple and accessible example: assume that $F_{\mu\nu}$ is (almost) periodic and oscillating within $P$, so our pulse is a finite wavetrain of $N$ cycles. (We will show later that our results hold for more realistic pulse shapes.) With these assumptions, the integrand of (\ref{FOURIER.MFRAC}) simplifies. One finds 
\be \label{FOURIER.MFRAC2}
  S_{fi} \!= \!\frac{1}{k_\p}\delta^3 (\bm{\mathsf{p}}' + \bm{\LCv{p}} -\bm{\LCv{k}}')\! \int\! \ud (k.x) \, e^{i(\bar{\LCv{y}}+\bar{\LCv{y}}'-{\LCv{x}}')k.x}{M}(k.x)\;,
\ee
where the lightfront momentum fractions in the exponent have become {\it quasi}-momentum fractions for the field in question\footnote{The quasi-momenta are not necessarily the same as for an IPW, even in a regular wavetrain, as they acquire transverse components. This does not alter the following arguments, though, so it is simplest to think of (\ref{QUASIMOM.INF}) as the definition of $q_\mu$ for this discussion.}, i.e. $\LCv{y}\equiv {p_+}/{k_+} \to \LCv{\bar{y}}\equiv {q_+}/{k_+} $ and $\LCv{\bar y}'$ similarly.  $M$ is periodic and oscillating.  Reducing the integration to a single period and squaring up, one finds
\be \label{SINSIN}
  \frac{|S_{fi}|^2}{VT} \sim \delta^3 (\bm{\mathsf{p}}' + \bm{\LCv{p}} -\bm{\LCv{k}}')\, \bigg|\ \frac{\sin N \pi \LCv{z}}{\sin \pi\LCv{z}}\ \sum_s M_s \, \text{sinc}\, \pi (\LCv{z} - s)\bigg|^2\; .
\ee
The $N$ cycles of the beam generate the ratio of sines, while the final sum is a Fourier expansion of the single cycle contribution. The sum in $|S_{fi}|^2$ is {\it coherent}, unlike that in (\ref{SSq}). Coherent sums are naturally associated with interference effects, which here are those of a diffraction pattern: the product of the $\sin/\sin$ factor and the $\text{sinc}$ function is nothing but the intensity distribution of $N$ slit diffraction. We thus expect a series of `peaks' and `troughs' in the pair production rate, and it is natural to ask what the peaks correspond to.

The important term is the ratio of sines. Its argument is $\LCv{z}\equiv \bar{\LCv{y}}+\bar{\LCv{y}}'-\LCv{x}'$.  This contains the longitudinal momentum transfer from the probe photon to the pair, as in (\ref{snew}), and also the {\it average} longitudinal momentum transfer from the laser to the pair (responsible for the transformation of $\LCv{y}$ to $\LCv{\bar y}$ etc). Hence, $\LCv{z}$ describes the combined longitudinal momentum transfer. It is clear that the ratio of sines is peaked when $\LCv{z}=n\in\mathbb{Z}$, which just says, multiplying through by $k_\p$, that we have peaks when the lightfront momentum transfer is a multiple of the laser frequency: the peaks are therefore a resonance effect. Combining this resonance condition with the delta functions conserving three-momentum, we see that the peaks have the kinematic support
\be \label{deltaspect}
	k'_\mu + n k_\mu = q_\mu + q_\mu' \;,
\ee
which looks just like an integer number of photons producing a heavy pair. Thus, the photon number and effective mass arise together as a consequence of a resonance effect. Note that when $\LCv{z}=n\in\mathbb{Z}$ is interpreted as a photon number, then from (\ref{FOURIER.MFRAC}) we see that this is conjugate to phase $k.x$, consistent with optics results.

\begin{figure}[t!]
\includegraphics[width=0.33\columnwidth]{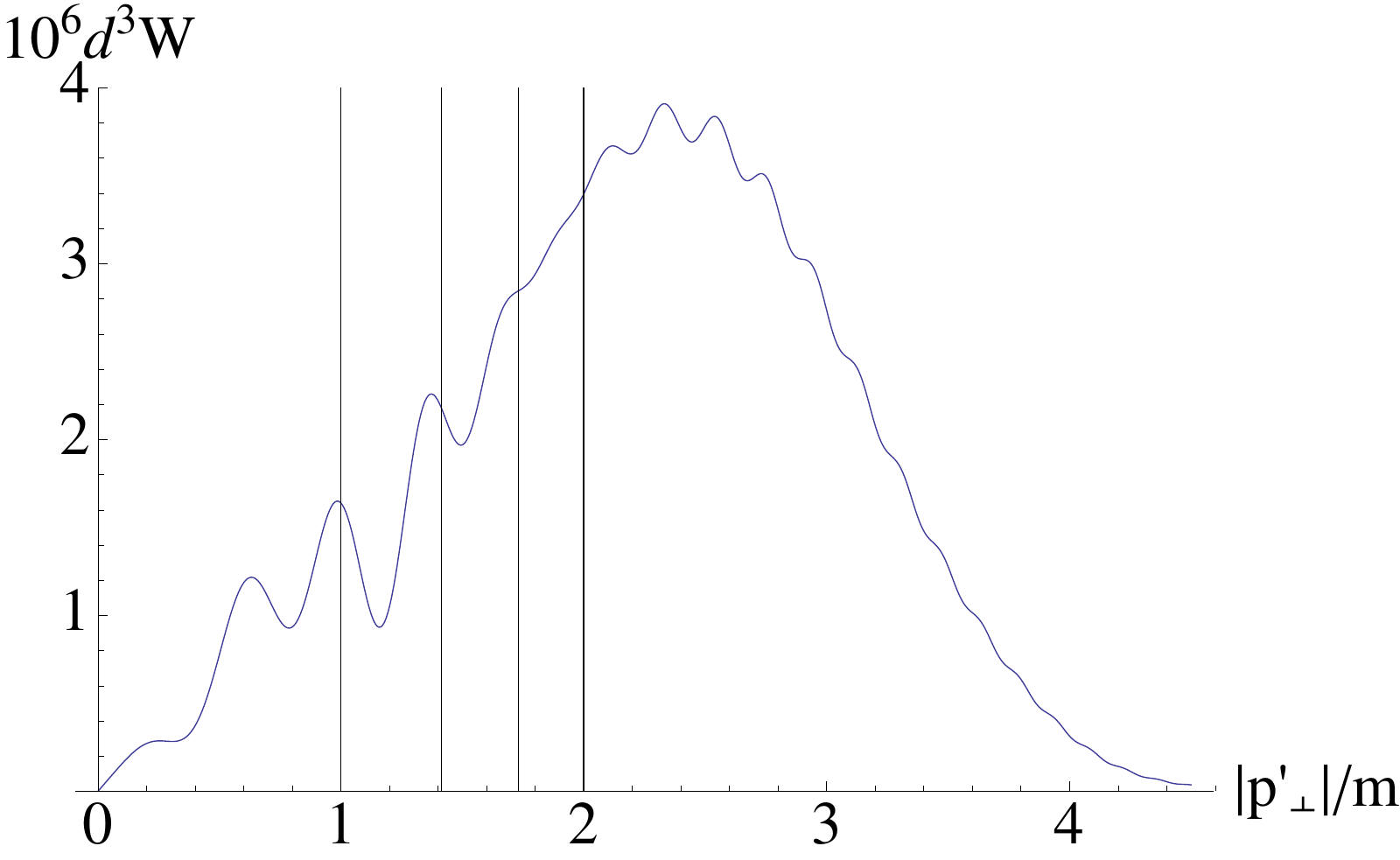}
\includegraphics[width=0.33\columnwidth]{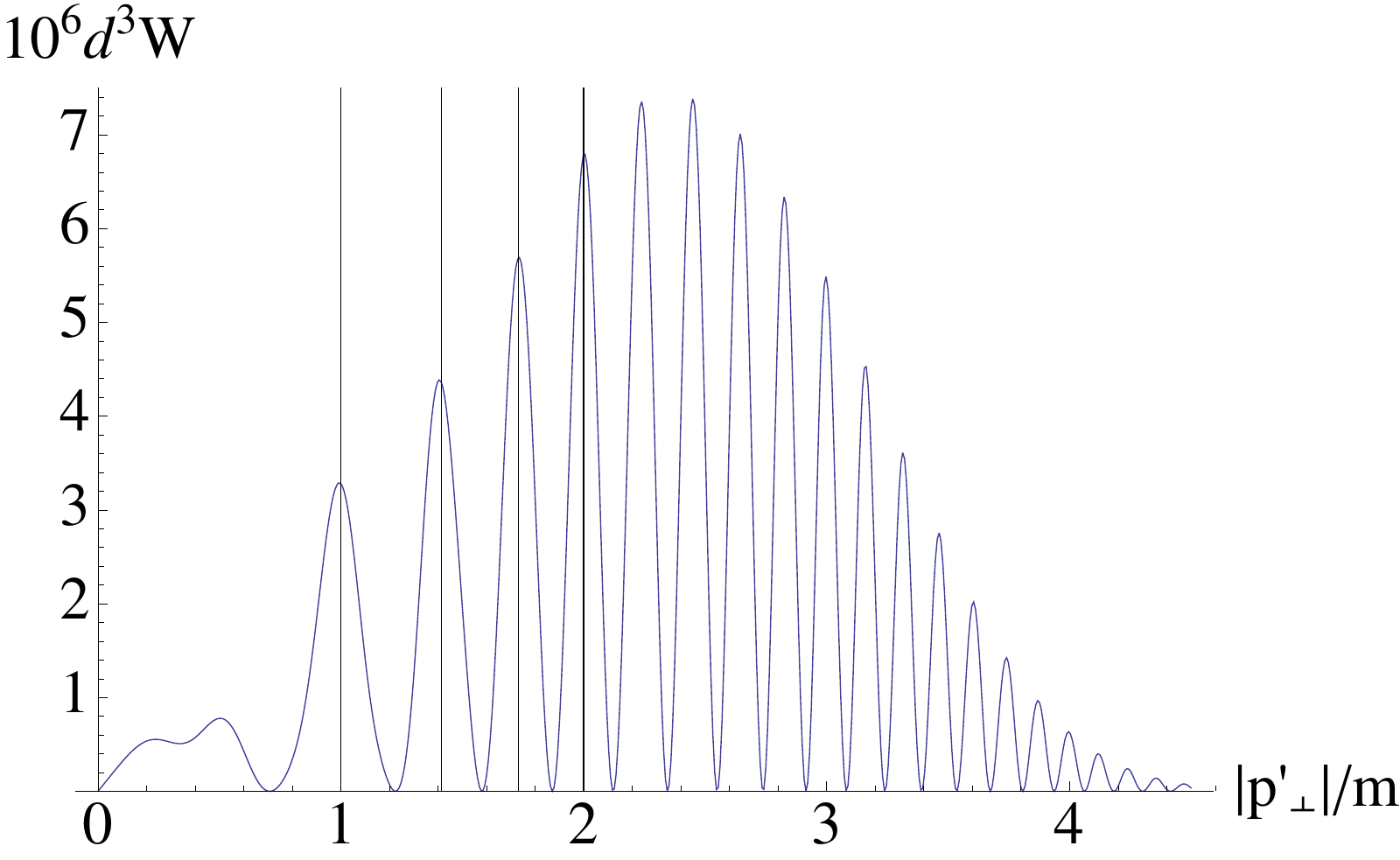}\includegraphics[width=0.33\columnwidth]{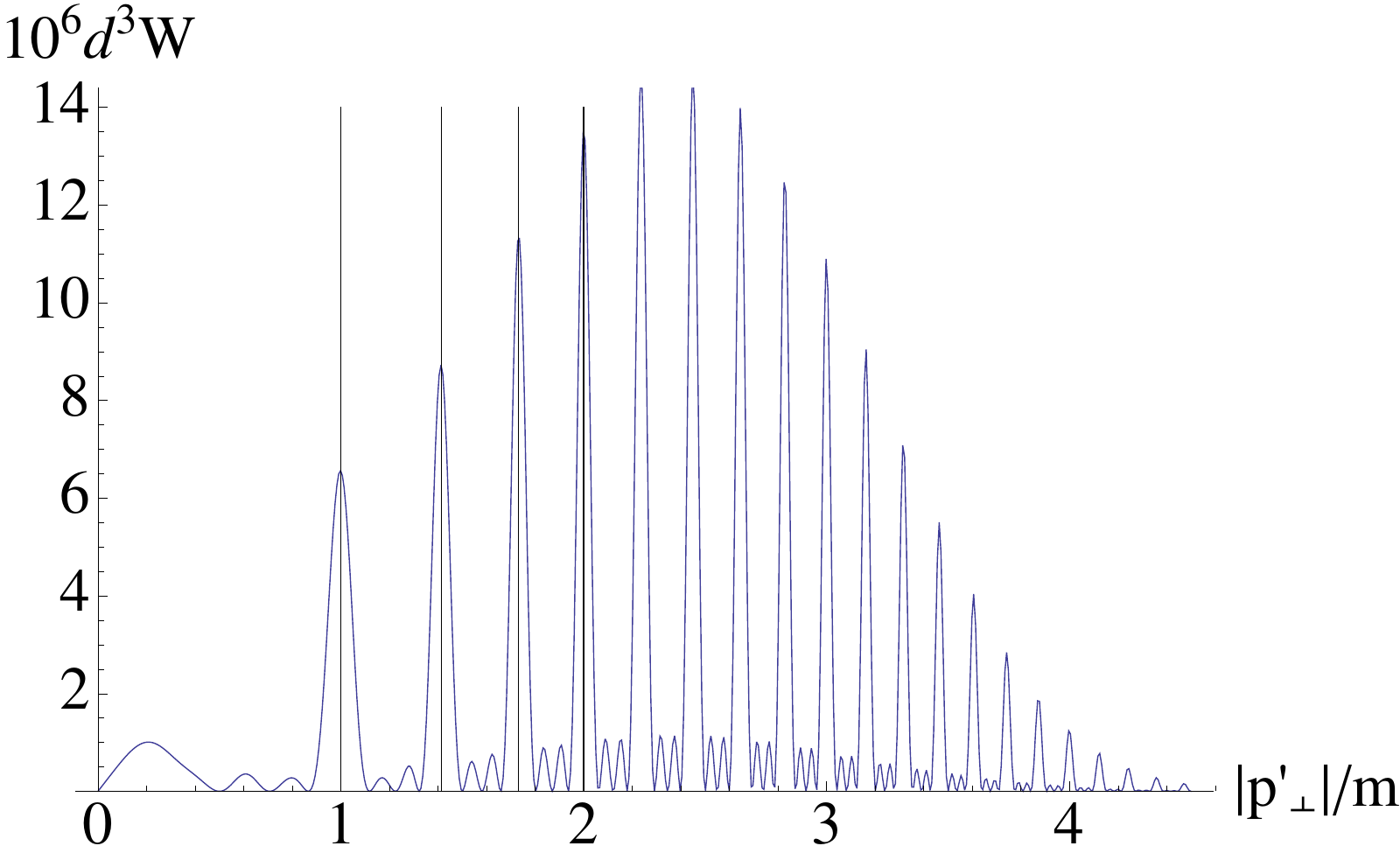}
\caption{\label{FIG:SECOND} Triple differential cross section [a.u.] as a function of transverse positron momentum, for $N=1$, 2 and 4 cycles. Fixed transverse angle $\phi=\pi$ and half maximum lightfront component $p'_\m=k'_\m/2$. Black (vertical) lines indicate the IPW result -- a delta comb centred on the resonances, corresponding to quasi-momentum conservation. }
\end{figure}

Let us examine the diffraction pattern. For simplicity we look at the triple differential cross section, with the background field $f_1(k.x)=\sin(k.x)$ for $0\leq k.x \leq 2\pi N$, zero otherwise, and $f_2=0$ for linear polarisation. The field strength has a step function edge, but it keeps things simple for now: we will shortly smooth everything out. The emission rate is plotted in Fig.~\ref{FIG:SECOND} for $N=1,2,4$ cycles of the laser (corresponding to $4$, $8$, $16$ fs pulse duration). The form of the rate changes significantly when we go above one cycle, which corresponds to the change from single to multiple slit diffraction: one can see the appearance of the $N-2$ subpeaks characteristic of such patterns.This is the origin of the oscillatory substructure observed recently in pair production rates \cite{Hebenstreit:2009km,MockenNew}.

As one takes the number of cycles to infinity, the interference terms drop out of $|S_{fi}|^2$ and the sum becomes incoherent; {\it only} the resonances in the spectrum survive in the IPW limit. The spectrum collapses to a delta comb supported on (\ref{deltaspect}), and the IPW results (\ref{SSq}), described entirely in terms of photon number and the effective mass, are correctly recovered. This is illustrated in Fig.~{\ref{FIG:SECOND}} by the vertical black lines, which show the locations of the resonances and represent the delta comb of the IPW limit: the peaks in the pulsed rate are centred on the IPW results, which dominate as $N\to\infty$. 

\subsection*{Example 2: smooth envelopes.}
\begin{figure}[t!]
\centering \includegraphics[width=0.3\textwidth]{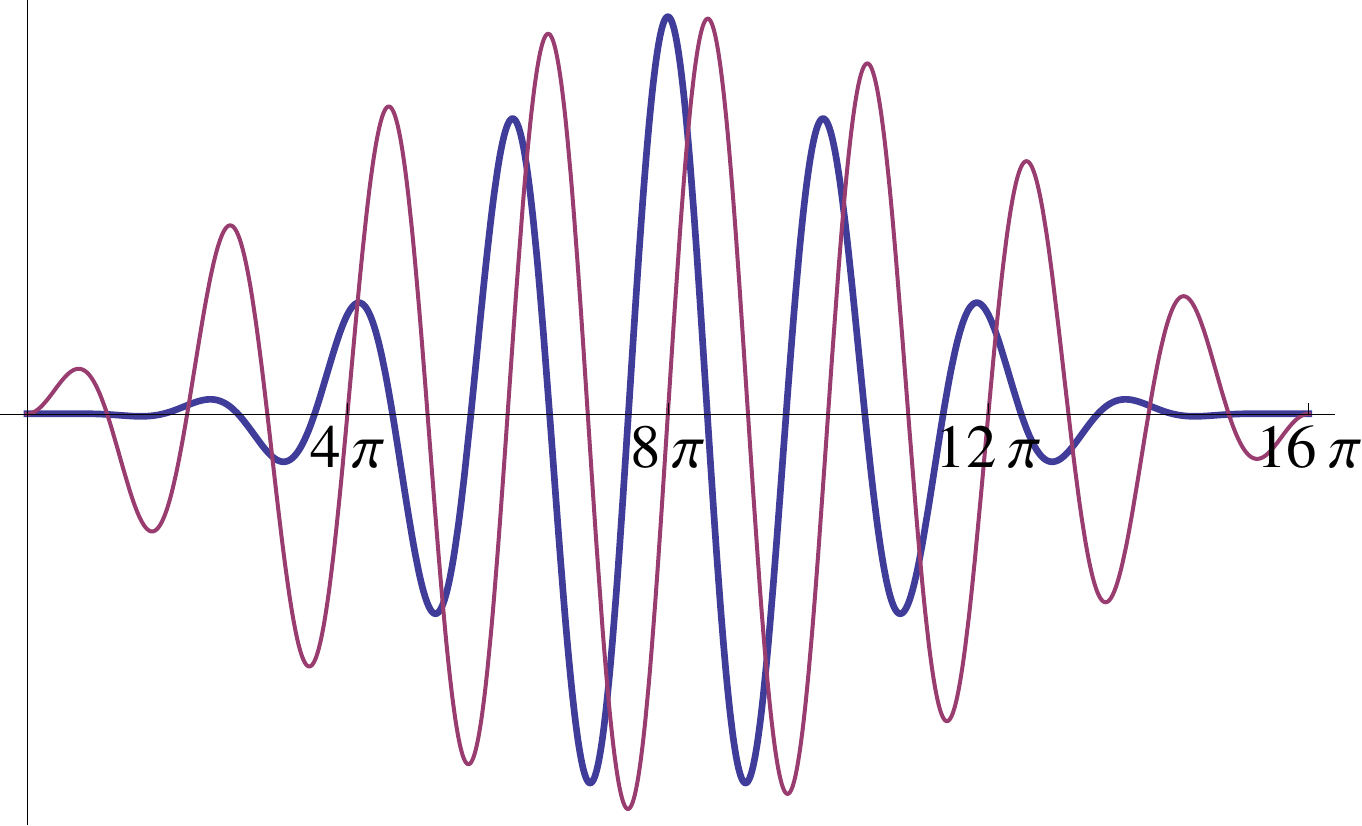}
\caption{\label{PulsePic} Two 8-cycle pulse profiles modulated by $\sin(x/16)$ (red, broad) and $\sin^4 x/16$ (blue, narrow). }
\end{figure}
\begin{figure}[b!]
\includegraphics[width=0.5\textwidth]{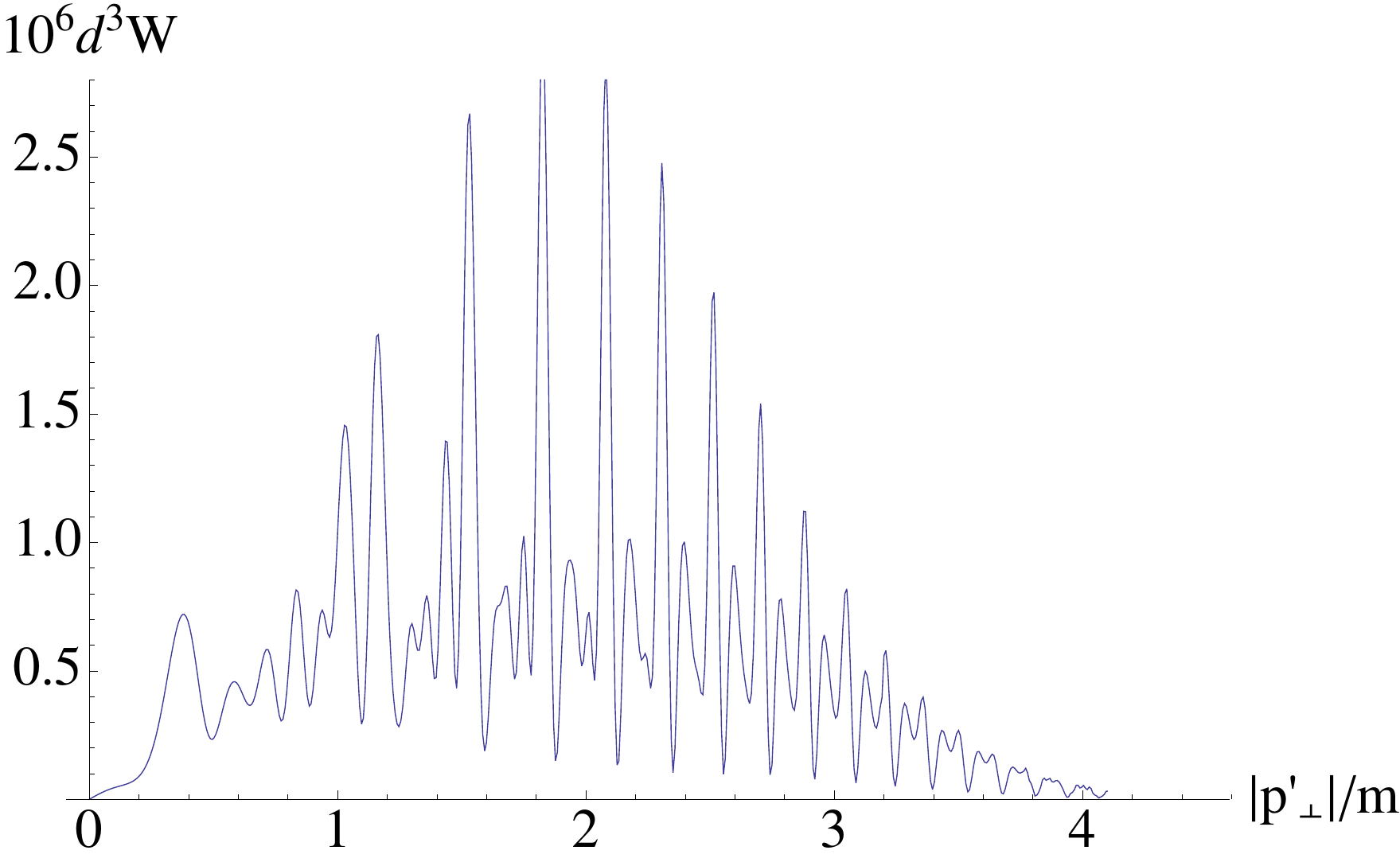}\includegraphics[width=0.5\textwidth]{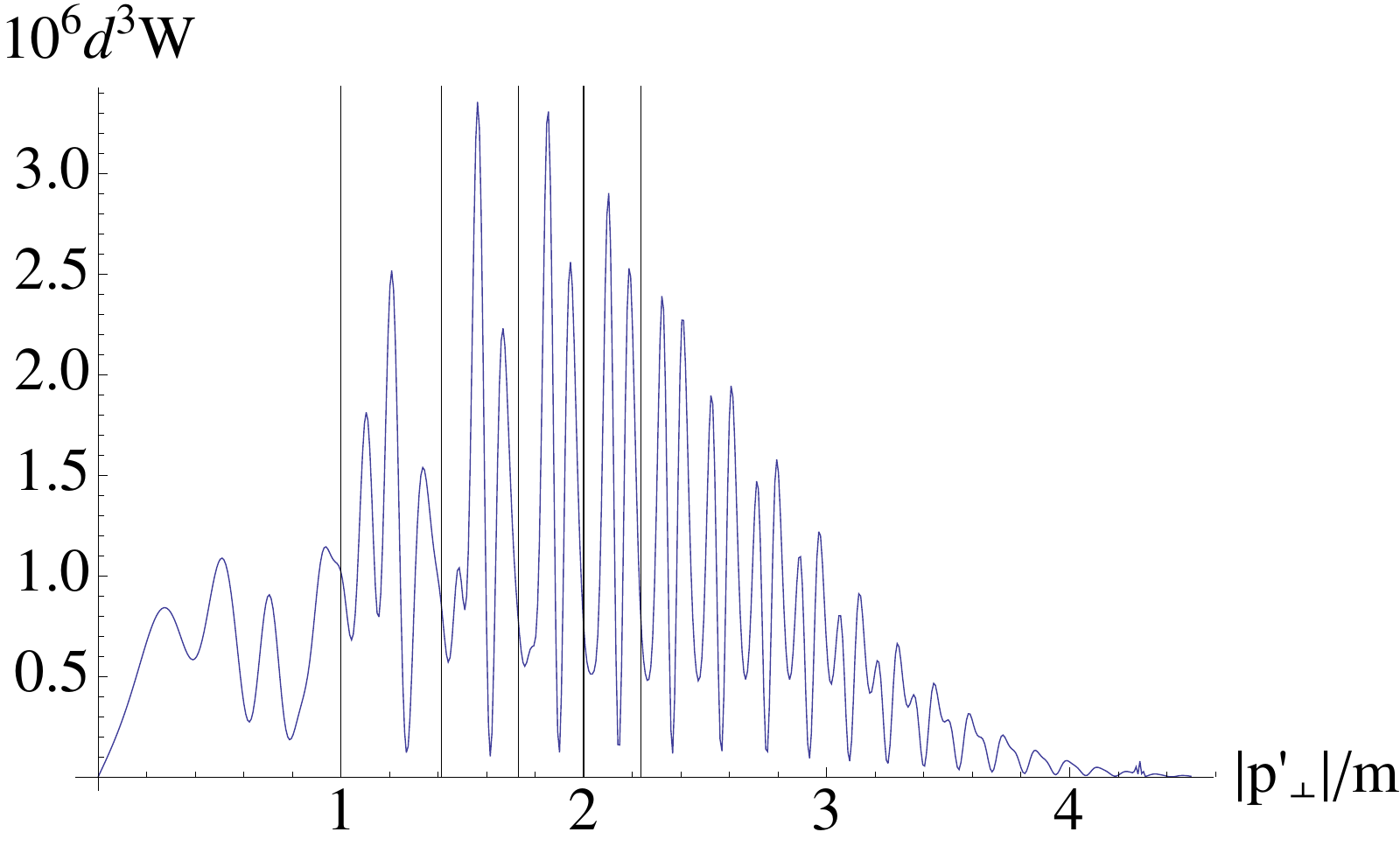}
\caption{\label{Smoothed} LEFT: 12-cycle (48 fs) pulse with $\sin^4$ profile. RIGHT: 8-cycle (32 fs) pulse with $\sin^2$ profile. The peaks do not match the IPW delta comb (vertical/black lines), indicating that the effective mass in a pulse differs from the infinite plane wave $m_*$.}
\end{figure}

It is important to ask what becomes of the interpretation above once we drop the assumption of periodicity and add an envelope function, such that the pulse profile looks something like that of Fig.~\ref{PulsePic}, and the field strength vanishes smoothly at the edges of the pulse. A nice model of such pulses was introduced in \cite{Mackenroth:2010jk} and studied in \cite{Heinzl:2010vg}: sticking to linear polarisation, one takes ($f_1$ or) $\dot f_1(k.x)=\sin^v(k.x/2N)\sin(k.x)$ for $0\leq k.x \leq 2\pi N$ and zero otherwise, where $v$ describes the envelope function: examples are shown in Fig.~\ref{PulsePic}.  Plotting the pair production rates reveals the same diffraction-like pattern as for the periodic model above -- the spectra feature a series of strong peaks, with a rich substructure, see Fig.~\ref{Smoothed}.   

Let us now compare these spectra with the IPW results in a little more detail. Adding an envelope seems to most strongly affect the low energy part of the pair spectrum. We note that a significant part of the spectrum lies below the momentum threshold for pair production in the IPW limit: on our plots this corresponds to the signal to the left of $|\vc{p}'_\LCperp|/m=1$ which is not allowed by conservation of {\it quasi}-momentum. The most interesting effect is that, as shown in the right hand panel of Fig.~\ref{Smoothed}, the strong peaks of the emission rate {\it do not} match those of the IPW limit.  The persistence of the diffraction pattern implies that some preferred momentum remains in the game, but it is not obvious how one can identify this since periodicity is lost: expressions for the S--matrix element do not explicitly reveal what average (c.f.\ the discussion preceding (\ref{QUASIMOM.INF})) the process is sensitive to.  Preliminary investigations into the distribution of the peaks, using an approximate numerical method to reconstruct the quasi-momentum, indicate that the corresponding effective electron mass is {\it lowered} in more realistic pulse geometries \cite{Heinzl:2010vg}.

\section{Conclusions}
We have given a clear and physical interpretation of the effective electron mass, and its role in pair production, by employing lightfront field theory. We have seen that pair production in a plane wave of finite temporal extent is a diffractive process: the emission spectrum can be interpreted as an interference pattern with a rich substructure. The rate exhibits resonant behaviour when the {\it laser averaged} lightfront momentum transfer to the pair is a multiple of the laser, or driving, frequency. For regular pulses with large numbers of cycles, this resonance condition recovers the IPW kinematics expressed purely in terms of the effective electron mass. Interestingly, the emission rates indicate that if an effective mass continues to play a role in pulsed pair production, it is lower than the $m_*$ of the infinite plane wave. The form, and phenomenological consequences of the effective mass will be further investigated in a future publication. Another important topic to be addressed is the development of systematic, controlled expansions which allow one to go beyond the plane wave approximation and tackle more realistic pulse geometries.  
\subsection*{Acknowledgements}
A.~I. thanks the organisers of LC2010 for inviting him to the conference, and James Vary for interesting discussions. M.~M. and A.~I. are supported by the European Research Council under Contract No. 204059-QPQV, and the Swedish Research Council under Contract No. 2007-4422.


\begin{thebibliography}{99}


\bibitem{Heinzl:2008an}
  T.~Heinzl and A.~Ilderton,
  Eur.\ Phys.\ J.\  D {\bf 55} (2009) 359
  
\bibitem{Bulanov:2004de}
  S.~S.~Bulanov, N.~B.~Narozhny, V.~D.~Mur and V.~S.~Popov,
  Phys.\ Lett.\  A {\bf 330} (2004) 1

\bibitem{Bulanov:2010ei}
  S.~S.~Bulanov {\it et al.}, 
  Phys.\ Rev.\ Lett.\  {\bf 104} (2010) 220404
  
\bibitem{Hebenstreit:2009km}
  F.~Hebenstreit, R.~Alkofer, G.~V.~Dunne and H.~Gies,
  Phys.\ Rev.\ Lett.\  {\bf 102}, 150404 (2009)

\bibitem{Dumlu:2010ua}
  C.~K.~Dumlu and G.~V.~Dunne,
  Phys.\ Rev.\ Lett.\  {\bf 104} (2010) 250402.

\bibitem{MockenNew}
  G.~R.~Mocken, M.~Ruf,  C.~M\"uller, C.~H.~Keitel,
  Phys. Rev. A {\bf 81} (2010) 022122.


\bibitem{Schutzhold:2008pz}
  R.~Sch\"utzhold, H.~Gies and G.~Dunne,
  Phys.\ Rev.\ Lett.\  {\bf 101}, 130404 (2008)
    
\bibitem{Ruf:2009zz}
  M.~Ruf {\it et al.}, 
  Phys.\ Rev.\ Lett.\  {\bf 102}, 080402 (2009)

\bibitem{Hallin:1994ad}
  J.~Hallin and P.~Liljenberg,
  Phys.\ Rev.\  D {\bf 52} (1995) 1150

\bibitem{MoortgatPick:2009zz}
  G.~Moortgat-Pick,
  J.\ Phys.\ Conf.\ Ser.\  {\bf 198} (2009) 012002.
  
\bibitem{Heinzl:2009nd}
  T.~Heinzl, D.~Seipt and B.~Kampfer,
  Phys.\ Rev.\  A {\bf 81} (2010) 022125.
  
\bibitem{Nikishov:1963}%
  A.~I.~Nikishov and V.~I.~Ritus, Zh.\ Eksp.\ Teor.\ Fiz.\ \textbf{46} , 776 (1963) 

\bibitem{Nikishov:1964a}%
  A.~I.~Nikishov and V.~I.~Ritus, Zh.\ Eksp.\ Teor.\ Fiz.\ \textbf{46}, 1768 (1964) 

\bibitem{Narozhnyi:1964}%
  N.~B.\ Narozhnyi, A.~Nikishov, and V.~Ritus,
  Zh.\ Eksp.\ Teor.\ Fiz.\ \textbf{47}, 930 (1964) 

\bibitem{Bamber:1999zt}
  C.~Bamber {\it et al.},
  Phys.\ Rev.\  D {\bf 60}, 092004 (1999).
  
\bibitem{Heinzl:2008rh}
  T.~Heinzl and A.~Ilderton,
  Opt.\ Commun.\  {\bf 282}, 1879 (2009)
  
\bibitem{Sengupta:1952}%
  N.~Sengupta, Bull.~Math.~Soc. (Calcutta), \textbf{44}, 175 (1952).
  
\bibitem{Heinzl:2003jy}
  T.~Heinzl,
  arXiv:hep-th/0310165.
  
\bibitem{Schwinger:1951nm}
  J.~S.~Schwinger,
  Phys.\ Rev.\  {\bf 82}, 664 (1951).

\bibitem{McD-talk}
  K.~T.~McDonald, in A.~K.~Das and T.~Ferbel, \textit{Probing Luminous And Dark Matter}, Proceedings of Symposium in honor of Adrian
  Melissinos, Rochester, 1999, World Scientific, Singapore (2000). 

\bibitem{volkov:1935}%
  D.~Volkov, Z.~Phys.\ \textbf{94}, 250 (1935).

\bibitem{Mackenroth:2010jk}
  F.~Mackenroth, A.~Di Piazza and C.~H.~Keitel,
  Phys.\ Rev.\ Lett.\  {\bf 105} (2010) 063903.

\bibitem{Heinzl:2010vg}
  T.~Heinzl, A.~Ilderton and M.~Marklund,
  Phys.\ Lett.\  B {\bf 692} (2010) 250.


\end{thebibliography}
\end{document}